\documentclass[final]{aipproc}
\layoutstyle{8x11double}

\newcommand{\gr}[1]{\mathrm{#1}}

\begin{document}

\title{Phase diagram of dense quark matter in QCD-like theories}

\author{Tom\'a\v{s} Brauner}{address={Faculty of Physics, University of Bielefeld, D-33501
Bielefeld, Germany}, altaddress={Department of Theoretical Physics, Nuclear
Physics Institute ASCR, CZ-25068 \v{R}e\v{z}, Czech Republic}}

\begin{abstract}
I report the results of a series of works on the phase diagram of theories with
a different number of colors and/or quarks in a different representation than
in QCD. Similarities as well as differences as compared to the real world are
pointed out, focusing in particular on the interplay of confinement and chiral
symmetry breaking. It will be argued that recent lattice data may provide us
with a clue to understand deconfinement in cold dense quark matter.
\end{abstract}

\maketitle

\section{Motivation}
Our knowledge about the region of the phase diagram of quantum chromodynamics
(QCD) at low temperatures and high densities is still rather rudimentary. The
reasons are twofold. First, standard lattice Monte Carlo techniques suffer from
the formidable sign problem at high densities. Second, available experimental
data from the astrophysics of compact stars are not constraining enough so far.

In this situation, it is worthwhile to investigate theories which do not
describe the real world, but still may give us some hint at the nonperturbative
physics of QCD at moderate densities. I will discuss a class of theories
(hereafter referred to as QCD-like) with quarks in a different representation
of the (possibly also different) color group with the common property that this
representation is (pseudo)real. Later, I will focus on two specific examples, namely
two-color QCD with fundamental quarks and (any-color) QCD with adjoint quarks
\cite{Kogut:2000ek}. However, the most striking features of these theories
follow directly from the reality of the quark representation.

The very first of them, which forms the basic motivation for their study, is
the fact that the QCD-like theories do not suffer from the sign problem at
nonzero baryon chemical potential.\footnote{In case of \emph{pseudoreal} quarks
an even number of flavors is necessary to avoid the sign problem. With an odd
number of flavors, the determinant of the Dirac operator is real, but may be
negative.} This makes lattice simulations at high density possible, and can
thus provide the much needed model-independent input on the equation of state
of cold dense quark matter.

Second, baryons in QCD-like theories are bosons as one can combine two quarks
to make a color singlet. This means that the low-density matter looks very much
different than in the real world. Indeed, finite density is accomplished by a
Bose--Einstein condensate (BEC) of bosonic baryons (diquarks) rather than by
the Fermi sea of nucleons. On the other hand, this brings important technical
advantages. One does not have to deal with three-body physics at low baryon
density. Also, at high density where quark matter is expected to deconfine,
Cooper pairing of quarks results in a gauge-invariant order parameter, making
dense matter a quark superfluid rather than a superconductor. Due to these
reasons, one has a decent chance to describe both low- and high-density matter
in QCD-like theories within a single theoretical (model) framework. In the real
world, this is something that nuclear astrophysics can only dream of.

\section{Two-color QCD}
In this section, I will discuss QCD with quarks in the fundamental
representation of the color $\gr{SU}(2)$ group. However, most of the
conclusions hold without change for quarks in any \emph{pseudoreal}
representation. (This class of theories was dubbed \emph{type-II} in
\cite{Zhang:2010kn}.) In this case, the wave function of a color-singlet
diquark is antisymmetric in color. Assuming spin-zero pairing which gives the
largest energy gain, the Pauli principle implies that the wave function must
also be antisymmetric in flavor. The baryon is therefore composed of two quarks
of different flavors.

Thanks to the (pseudo)reality of the quark representation, the quark field has
the same color transformation properties as its charge conjugate. It is then
advantageous to trade the right-handed component of the quark (Dirac) spinor
for the charge-conjugated left-handed quark. Instead of $N_{\rm f}$ flavors of
Dirac fermions one then in effect deals with $2N_{\rm f}$ flavors of Weyl
fermions. Consequently, the global flavor symmetry of the QCD-like theory in
the chiral limit is $\gr{SU}(2N_{\rm f})$ rather than the usual
$\gr{SU}(N)_{\rm L}\times\gr{SU}(N)_{\rm R}\times\gr{U}(1)_{\rm B}$. Note that
the baryon number $\gr{U}(1)_{\rm B}$ is embedded in the simple group
$\gr{SU}(2N_{\rm f})$.

The extended symmetry of course affects the physical spectrum of the theory.
Put in simple terms, the invariance under the exchange of right-handed quarks
and left-handed antiquarks implies that multiplets in the spectrum contain
states with different baryon number. In particular, we will find mesons and
baryons (diquarks) in the same multiplet. The ground state of two-color QCD
exhibits, very much like real QCD, a chiral condensate which breaks the flavor
symmetry as $\gr{SU}(2N_{\rm f})\to\gr{Sp}(2N_{\rm f})$. There are $2N_{\rm
f}^2-N_{\rm f}-1$ Nambu--Goldstone (NG) bosons. In the case $N_{\rm f}=2$, to
which I will from now on restrict, this means altogether five NG bosons. Three
of them form the usual isovector of pions, while the remaining two are a
diquark and an antidiquark.

\begin{figure}
\includegraphics[width=0.45\textwidth]{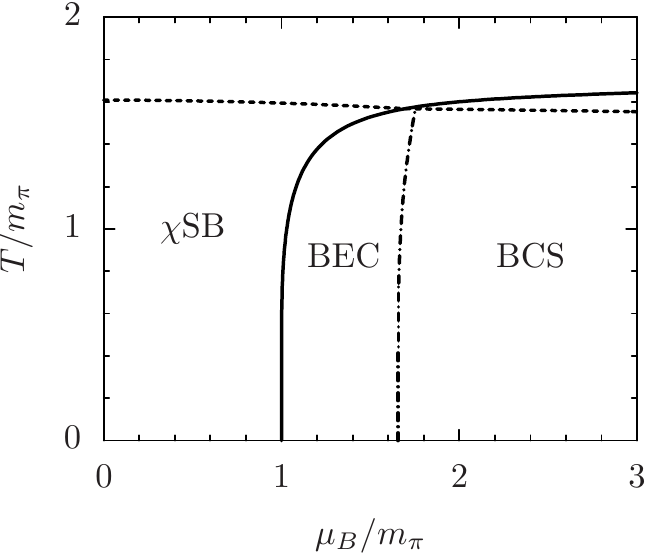}
\caption{Phase diagram of two-color QCD with two light quark flavors. The labels distinguish various
regions with qualitatively different physical behavior. In ``$\chi$SB'' only the chiral
condensate is nonzero; this is analogous to the hadronic phase of QCD. ``BEC''
labels a phase in which the system behaves as a Bose--Einstein condensed gas of
tightly bound diquarks. In ``BCS'', the physics is dominated by a Fermi sea of
quarks, slightly distorted by loose quark pairing. See the text for explanation of the lines.}
\label{Fig:2cPD}
\end{figure}
Even when the flavor symmetry is broken explicitly by (small) quark masses,
there will still be light pseudo-NG bosons. The fact that some of them carry
baryon number has  far-reaching consequences: it puts the finite-density part
of the phase diagram in reach of chiral perturbation theory
\cite{Kogut:2000ek}. The result of a model calculation of the phase diagram of
two-color QCD with two light quark flavors \cite{Brauner:2009gu} is shown in Fig.~\ref{Fig:2cPD}.
Chiral perturbation theory successfully describes the ``$\chi$PT'' and ``BEC''
phases as well as the second-order phase transition (solid line) separating
them. As the density increases, the diquarks get closely packed and the
relevant degrees of freedom become the quarks themselves. The superfluid ground
state is then well captured by the Bardeen--Cooper--Schrieffer (BCS) pairing
of quarks around the Fermi sea. The position of the smooth (BEC--BCS) crossover between the two
regimes can be indicated for example by a change of sign of the (nonrelativistic) quark
chemical potential and is shown in Fig.~\ref{Fig:2cPD} by the dash-dotted line.

In order to describe the BCS regime, one needs a model with quark degrees of
freedom such as that of Nambu and Jona-Lasinio (NJL)
\cite{Ratti:2004ra,Sun:2007fc}. In a version of the model augmented with the
Polyakov loop \cite{Brauner:2009gu}, one can simultaneously study the
deconfinement phase transition. In presence of dynamical quarks, this becomes a
smooth crossover, and is indicated in Fig.~\ref{Fig:2cPD} by the dashed line
(defined as a point at which the expectation value of the Polyakov loop is
$0.5$). We observe that the deconfinement temperature is essentially
insensitive to the chemical potential. Even though this is an obvious artifact
of the PNJL model, it can still be close to the actual behavior of two-color
QCD, as supported by recent lattice simulations \cite{Hands:2010gd}.

More difficult to understand seems the thermodynamic behavior of two-color
matter around the second-order BEC phase transition. Lattice data for pressure,
baryon density and energy density normalized to the (free quark gas)
Stefan--Boltzmann limits indicate peaks in all three quantities around the BEC
transition. This itself is qualitatively easy to understand already in the
chiral perturbation theory. However, the peak in the energy density turns out
to be an order of magnitude higher than in the other two observables. This
behavior is hard to reproduce in any model with just quark or diquark degrees
of freedom based on the global flavor symmetry \cite{Andersen:2010vu}.

\section{Adjoint QCD}
In this section, I will use QCD with adjoint quarks (of two or three colors) as
a specific example. However, most of what follows holds for quarks in any
\emph{real} representation. (This class of theories was called \emph{type-I} in
\cite{Zhang:2010kn}.) The important distinguishing property of this class of
theories is that the center symmetry associated with the deconfinement
transition is not broken by the presence of dynamical quarks. As a consequence,
in the chiral limit there are still two sharp phase transitions and it is a
well defined question whether these transitions coincide or not (unlike real
QCD where it is somewhat moot).

As a matter of fact, it has been known for a decade that in adjoint
(three-color) QCD, the chiral restoration temperature is much higher than that
of deconfinement \cite{Karsch:1998qj,Engels:2005te}. (The ratio of the
temperatures is about $7.8$.) Unlike the case of type-II theories, here the
diquark wave function is symmetric in color. This in turn changes the flavor
structure of the wave function; it must be symmetric as well, which opens up
the possibility of pairing of quarks of the same flavor.

The fact that deconfinement (or rather center symmetry breaking) is a sharp
phase transition also means that type-I theories exhibit interesting critical
behavior. In particular in two-color adjoint QCD deconfinement is a
second-order transition and we expect a tetracritical point in the phase
diagram where the deconfinement and BEC transition lines intersect
\cite{Sannino:2004ix}. In three-color adjoint QCD deconfinement is a
first-order transition so that the tetracritical point is replaced with two
nearby tricritical points where the second-order BEC line meets the first-order
deconfinement line \cite{Zhang:2010kn}.

The global flavor symmetry is the same as in the previous case, but the chiral
condensate breaks it in a different way, $\gr{SU}(2N_{\rm f})\to\gr{SO}(2N_{\rm
f})$. This leads in general to $2N_{\rm f}^2+N_{\rm f}-1$ NG bosons. In the
$N_{\rm f}=2$ case, there is still the isovector of pions, and in addition the
isovectors of diquarks and antidiquarks. The fact that diquark is an isovector
means that once it condenses at sufficiently high baryon chemical potential,
the isospin symmetry will be spontaneously broken. This is a rather rare
example of a spontaneous breaking of an exact vector symmetry, which is ruled
out in the vacuum by the Vafa--Witten theorem.

Using an improved version of the PNJL model with gauge sector given by a spin
model with a nearest-neighbor interaction \cite{Abuki:2009dt} allows one to
evaluate the expectation values of Polyakov loops in different representations
and thus to study the hypothesis of Casimir scaling \cite{Ambjorn:1984mb}. This
conjectures that the expression $\langle\ell_{\mathcal R}\rangle^{1/C_2(\mathcal R)}$,
where $C_2(\mathcal R)$ is the quadratic Casimir invariant, should be
independent of the representation $\mathcal R$. For two colors, one can obtain
the expectation value of the Polyakov loop in an arbitrary representation
(labeled by the ``spin'' $j$) analytically as
$\langle\ell_j\rangle=I_{2j+1}(2\alpha)/I_1(2\alpha)$, where $\alpha$ is a mean
field that depends implicitly on temperature and chemical potential
\cite{Zhang:2010kn}. In the high-temperature (high-$\alpha$) expansion, one
obtains
\begin{equation}
\langle\ell_j\rangle^{1/j(j+1)}=1-\frac1\alpha+\frac1{4\alpha^2}+
\frac{8j(j+1)-1}{96\alpha^3}+\cdots,
\label{casimir}
\end{equation}
which shows that the Casimir scaling is indeed very well satisfied; it is only
violated at the fourth order of the expansion. For three colors there is no
closed analytic expression for the expectation values of the Polyakov loops in
all representations. Nevertheless, Casimir scaling can still be studied
numerically \cite{Zhang:2010kn,Abuki:2009dt}. The importance of the result such
as Eq.~\eqref{casimir} is that it holds regardless of the actual value of
temperature and chemical potential and thus allows one to study Casimir scaling
even at finite density.

\section{Outlook and challenges}
The feasibility of lattice simulations at high density makes the QCD-like
theories very interesting  toy models for understanding nonperturbative QCD
physics in cold dense matter. At the moment, there are two main issues to be
properly understood. First is the peculiar thermodynamic behavior around the
BEC transition \cite{Hands:2010gd}. While this certainly does not describe the
real world, it is important for establishing a reliable overall agreement
between the results from analytic calculations and lattice simulations. Second,
recent lattice data hint at the possibility of a deconfinement transition at
low temperature and high chemical potential. Understanding this behavior would
provide a much needed input for the construction of the equation of state of
dense quark matter. It is therefore certainly worth further investigation.

\begin{theacknowledgments}
I gratefully acknowledge the collaboration of J.~O.~Andersen, K.~Fukushima,
Y.~Hidaka, D.~H.~Rischke, and T.~Zhang that gave rise to the results presented
here. This research was supported in part by the ExtreMe Matter Institute EMMI
in the framework of the Helmholtz Alliance Program of the Helmholtz Association
(HA216/EMMI), and by the Sofja Kovalevskaja program of the Alexander von Humboldt Foundation.
\end{theacknowledgments}

\bibliographystyle{aipproc}
\bibliography{references}

\end{document}